\documentclass[12pt]{iopart}

\usepackage{graphicx}
\usepackage{bm}
\usepackage{latexsym}
\usepackage{amssymb}

\begin{document}

\title[Intrinsic dynamics of the heart regulatory systems]{Intrinsic dynamics
of heart regulatory systems on short time-scales: from experiment to modelling}
\author{I.~A. Khovanov, N.~A. Khovanova, P.~V.~E. McClintock\\ and A. Stefanovska}
\address{Department of Physics, Lancaster University, Lancaster LA1 4YB, UK. }
\ead{i.khovanov@lancaster.ac.uk}


\begin{abstract}
We discuss open problems related to the stochastic modeling of cardiac
function. The work is based on an experimental investigation of the dynamics of
heart rate variability (HRV) in the absence of respiratory perturbations. We
consider first the cardiac control system on short time scales via an analysis
of HRV within the framework of a random walk approach. Our experiments show
that HRV on timescales of less than a minute takes the form of free diffusion,
close to Brownian motion, which can be described as a non-stationary process
with stationary increments. Secondly, we consider the inverse problem of
modeling the state of the control system so as to reproduce the experimentally
observed HRV statistics of. We discuss some simple toy models and identify open
problems for the modelling of heart dynamics.
\end{abstract}


\vspace{2pc}



\maketitle

\section{Introduction}

The human heart does not beat at a constant rate, even for a
subject in repose. Rather, there is strong variability of the
heart rate. The complexity of this heart rate variability (HRV)
presents a major challenge that has attracted continuing
attention. Many of the explanations proposed are by analogy with
paradigms used in physics to describe complexity, including:
deterministic chaos \cite{Ott:93}; the statistical theory of
turbulence \cite{Frisch:95}; fractal Brownian motion
\cite{Mandelbrot:82}; and  critical phenomenon \cite{Jensen:98}.
They have led to new approaches and time-series analysis
techniques including a variety of entropies
\cite{Kurths:95,Pincus:91,Costa:02}, dimensional analysis
\cite{Raab:06}, the correlation of local energy fluctuations on
different scales \cite{Kiyono:05}, the analysis of long range
correlation \cite{Peng:95}, spectral scaling
\cite{Hausdorff:96,Pilgram:99}, the multiscale time asymmetry
index \cite{Costa:05}, multifractal cascades
\cite{Ivanov:99,Ivanov:01}. All these measures allow one to
describe HRV as a non-stationary, irregular, complex fluctuating
process. Depending on the technique in use there has been a very
wide range of conclusions about the regulatory mechanism of heart
rate, ranging from a stochastic feedback configuration
\cite{Ivanov:98b} to the physical system being in a critical state
\cite{Kiyono:05}. HRV can also be considered in terms of the
interactions between coupled oscillators of widely differing
frequencies \cite{Stefanovska:99a}.

Although we now have this huge variety of tools and approaches
for the analysis of HRV, only the last-mentioned has enabled us
to understand the origins of some of the time-scales embedded
in HRV. Each time-scale (frequency) in the coupled oscillator
model \cite{Stefanovska:99a} is represented by a separate
self-oscillator that interacts with the others, and each of the
oscillators represents a particular physiological function. The
frequency variations in HRV can therefore be attributed to the
effects of respiration ($\sim$0.25\,Hz), and myogenic
($\sim$0.1\,Hz), neurogenic ($\sim$0.03\,Hz) and endothelial
($\sim$0.01\,Hz) activity. HRV also contains a fast (short
time-scale) noisy component which forms a noise background in
the HRV spectrum and can be modelled as a white noise source
\cite{Stefanovska:99a}. Its properties are currently an open
question, and one that is important for both understanding and
modelling HRV. A practical difficulty in experimental
investigations is the presence of a strong perturbation,
respiration, that occurs continuously and exerts a particularly
strong influence in modulating the heart rate. This modulation
involves several  mechanisms: via mechanical movements of the
chest, chemo-reflex, and couplings to neuronal control centres
\cite{Eckberg:03}. Spontaneous respiration gives rise to a
complex non-periodic signal, and this complexity is inevitably
reflected in HRV \cite{West:05r}. So, in order to understand
the properties of the fast noise, one would ideally remove the
respiratory perturbation and consider the residual HRV which
would then reflect fluctuations of the intrinsic dynamics of
the heart control system.

\begin{figure}[th]
\centerline {\includegraphics[width=14cm]{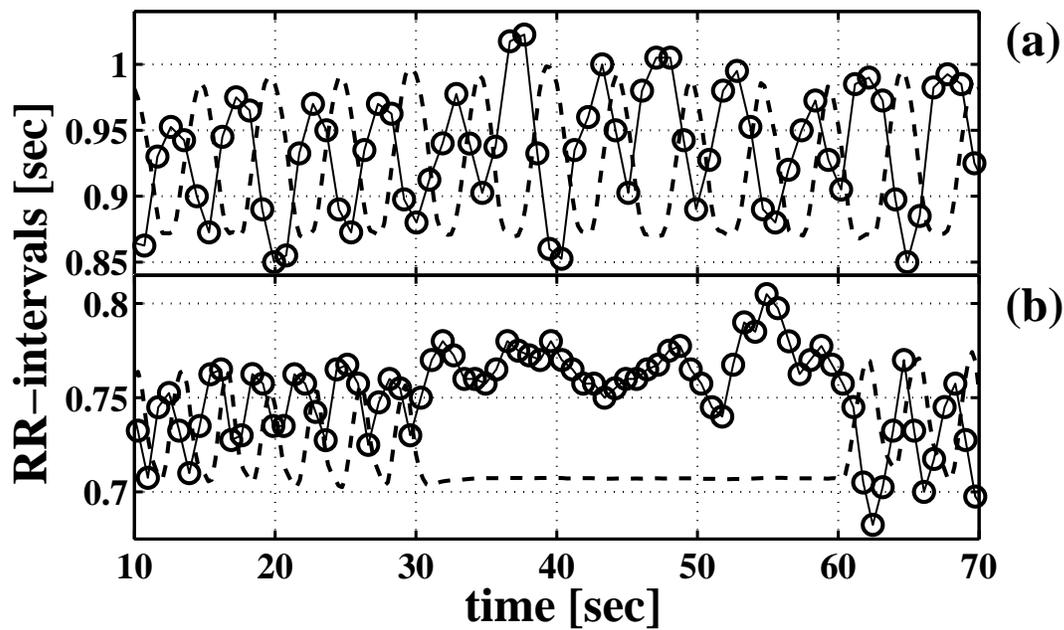}}
\caption{\label{fig1}  $RR$-intervals for
(a) normal (spontaneous) and  (b) intermittent respirations.
Respiration signals (arbitrary units) are shown by dashed lines.
 }
\end{figure}

\begin{figure}[ht]
\centerline{\includegraphics[width=12cm]{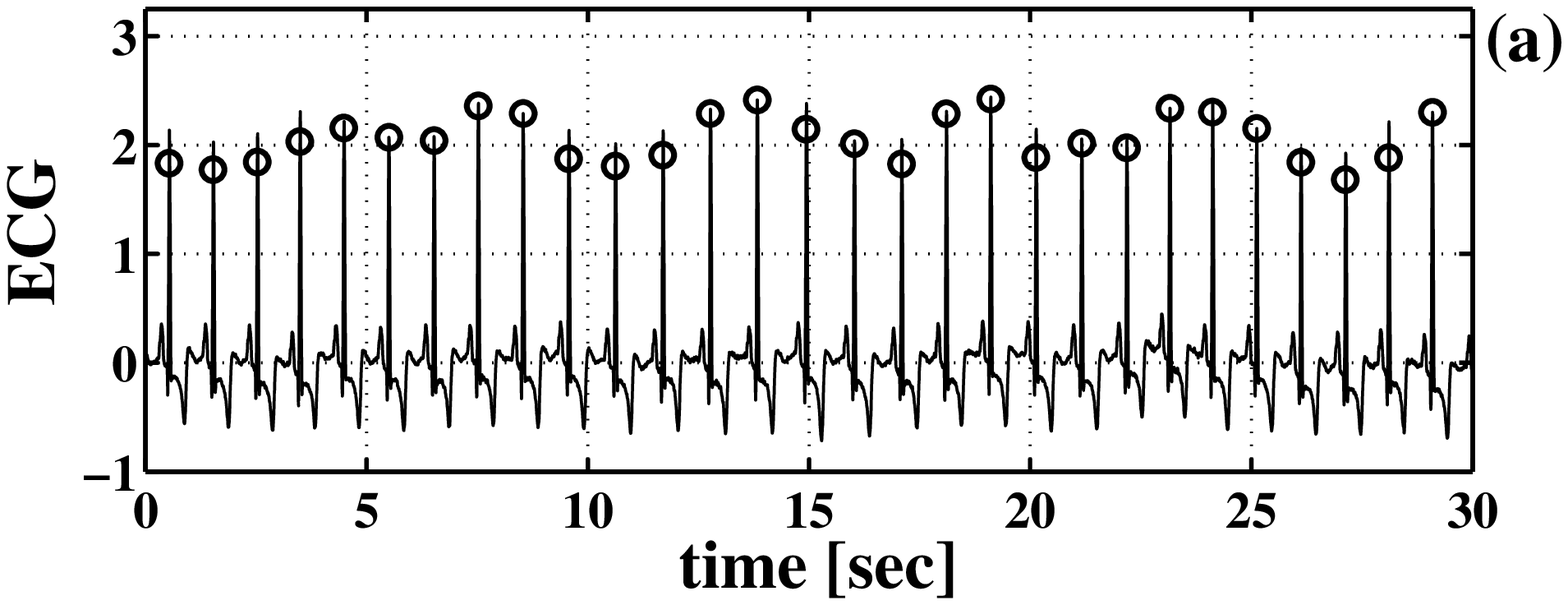}}
\vspace{0.2cm}
\centerline{\includegraphics[width=12cm]{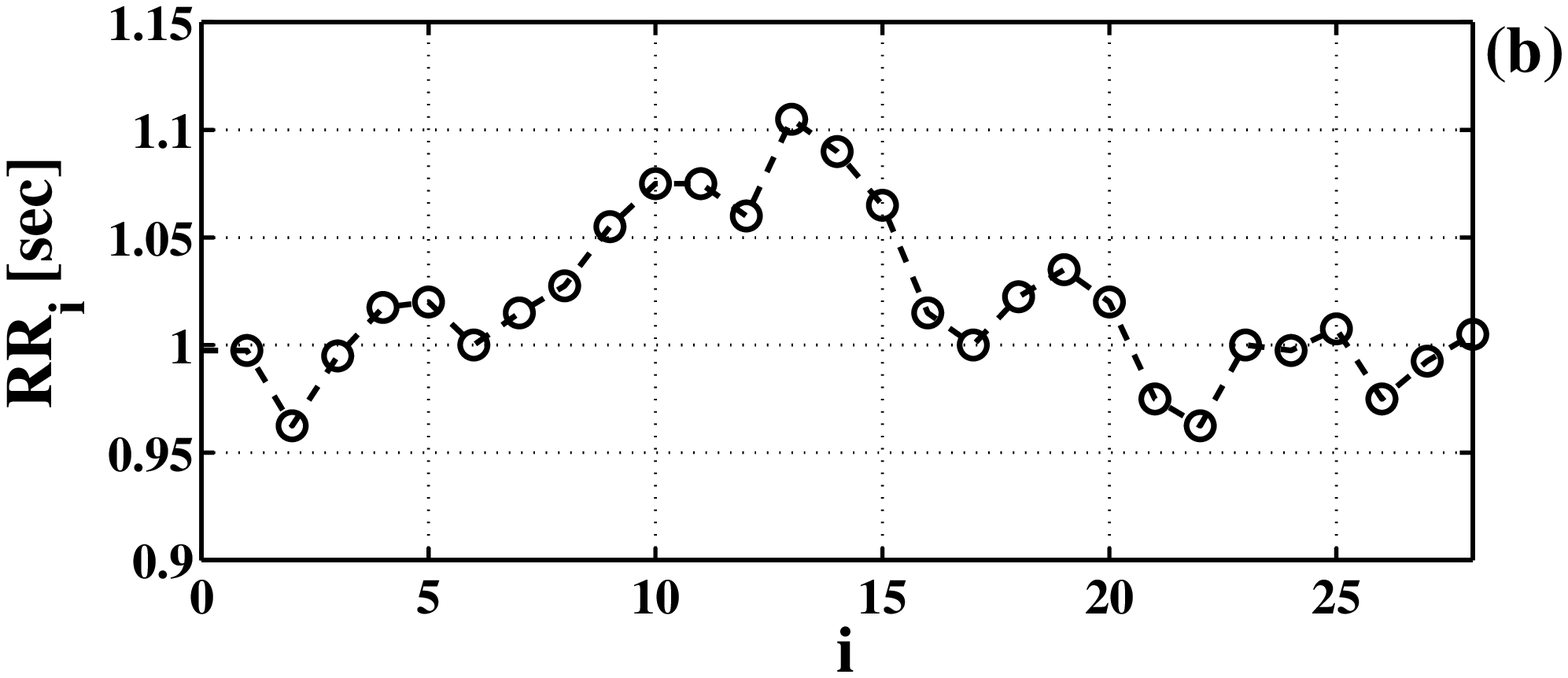}}
\caption{\label{fig2} (a) An ECG signal and (b) the
corresponding HRV ($RR$ intervals) signal. In (a) the R-peaks
are marked by $\opencircle$; the ECG signal is shown in
arbitrary units. }
\end{figure}

Consideration of the intrinsic activity of the heart control
system on short-time scales is important for general
understanding of the function of the cardio-vascular system,
leads potentially  to diagnostics of causes of arrhythmia
involving problems with neuronal control \cite{Doessel:06}, and
can be a benchmark for modeling  HRV. In this paper we present
the results of an experimental study of the intrinsic dynamics
of the heart regulatory system and discuss these results in the
context of modelling the fast noise component. A number of open
problems are identified.

\section{Experimental results}

We analyse the dynamics of the control system in the absence of
explicit perturbations by  temporarily {\it removing} the
continuing perturbations caused by respiration
[figure~\ref{fig1}(b)]. To do so, we perform experiments
involving modest breath-holding (apn{\oe}a) intervals. Note
that during long breath-holding the normal state of the
cardiovascular system is significantly modified
\cite{Parkes:06}. The idea of the experiments came from the
observation that spontaneous apn{\oe}a occurs during repose.
Apn{\oe}a intervals of up to 30 sec were used, enabling us to
avoid either anoxia or hyper-ventilation \cite{Parkes:06}.

Respiration-free intervals were produced by {\it intermittent
respiration}, involving an alternation between several normal
(non-deep) breaths and then a breath-hold following the last
expiration, as indicated by the dashed line in
figure~\ref{fig1}(b). The respiratory amplitude was kept close
to normal to avoid hyper-ventilation, and there were relatively
long intervals of apn{\oe}a when the heart dynamics was not
perturbed by respiration. It is precisely these intervals that
are our main object of analysis. The durations of both
respiration and apn{\oe}a intervals were fixed at 30 sec.

Measurements were carried out for 5 relaxed supine subjects,
and they  were approved by Research Ethics Committee of
Lancaster University. Note that the measurements presented have
been selected from a larger number of measurements to form a
set recorded under almost identical conditions of time and
duration, with the subjects avoiding either coffee or a meal
for at least 2 hours beforehand. They were 4 males and 1
female, aged in the range 29--36 years, non-smokers, without
any history of heart disease. We stress that the aim of the
current investigation was exploratory: to study typical
behaviour of the internal regulatory system; we have not
performed a large-scale trial of the kind widely used in
medicine when a large number of subjects is necessary because
of the need for subsequent statistical analysis of the data.
The electrocardiogram (ECG) and respiration signals were
recorded \cite{Stefanovska:99a} over 45-60 minutes. The ECG
signals were transformed to HRV by using the marked events
method for extraction of the $RR$-intervals which are shown in
figure~\ref{fig2}.

Figure~\ref{fig1} shows  $RR$-intervals found for the different types of
respiration. It is evident that respiration changes the heart rhythm very
significantly. Immediately after exhalation (b), there is an apn{\oe}a interval
where the heart rhythm fluctuates around some level. These fluctuations
correspond to the intrinsic dynamics of the heart control system. It is clear
from (a) that heart rate is {\it continuously} perturbed during normal
respiration, whereas in (b) one can distinguish an interval of intrinsic
dynamics corresponding to apn{\oe}a. Thus, the $j$th  interval of
apn{\oe}a  is characterized by the time series $\{RR_i\}$; here $i=1,2\ldots$
labels the $i$th $RR$-interval. Finally, we form a set $\{RR_i\}^j$ for
analyses by considering the set as  realizations of a random walk and analyzing
their dynamical properties as such.

\begin{figure}[t]
\centerline{ \includegraphics[width=14cm]{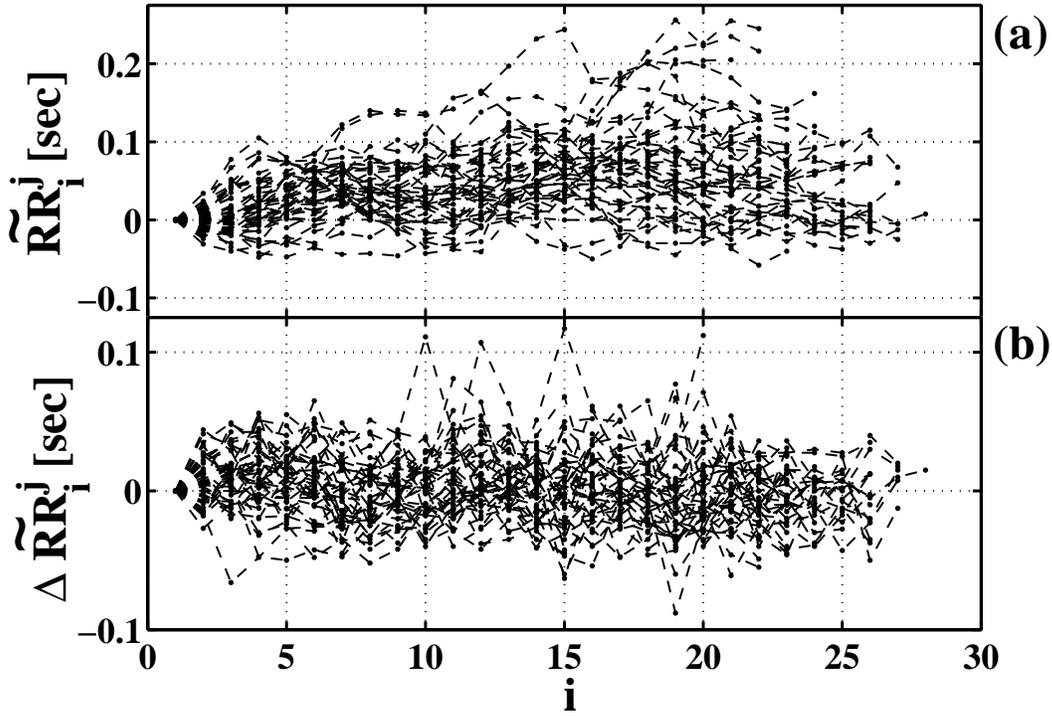}}
\caption{\label{fig3} (a) $RR$-intervals and (b) increments
$\Delta RR$ corresponding to apn{\oe}a intervals are shown. For
convenience of presentation, the difference between a given
value and the first value of each $j$th apn{\oe}a interval is
drawn in each case: $\widetilde {RR}_i^j=RR_i^j-RR_1^j$ and
$\Delta \widetilde {RR}_i^j=\Delta RR_i^j-\Delta RR_1^j$. }
\end{figure}

\begin{figure}[ht]
\centerline{ \includegraphics[width=14cm]{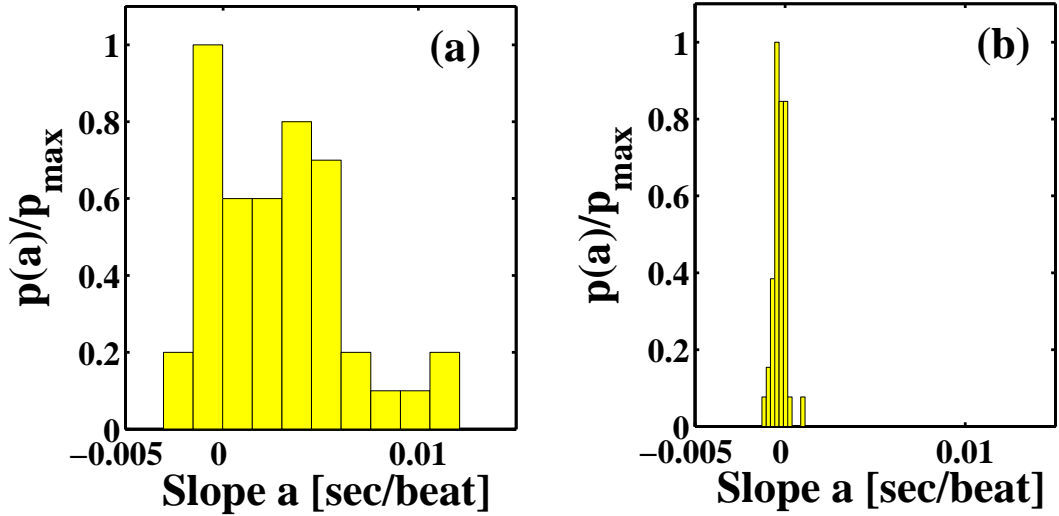}}
\caption{\label{fig4}  Distributions of trend slopes $P(a)$ of
the sets (a) $\{RR_i\}^j$ and (b) $\{\Delta RR_i\}^j$. }
\end{figure}

To reveal dynamics additional to $RR$-intervals, the
differential increments $\Delta RR_i=RR_{i+1}-RR_{i}$ were
analyzed. The differences between $RR$-intervals and their
increments are illustrated in figure~\ref{fig3}. Each apn{\oe}a
time-series $\{RR_i\}^j$ exhibits a trend that is describable
by the slope $a$ of a linear function $RR^j_i \propto a_j \, i$,
where $i$ is a heart beat number and $j$ marks $j$th apn{\oe}a
interval. The trend can be characterized by the distribution of
slopes $P(a)$ shown in figure~\ref{fig4} (a). For all
measurements the distributions are broad and their mean values
differ from zero. Thus the non-stationary nature of HRV on
short time-scales is clearly apparent. Note, that the
distributions $p(a)$ for the increments $\Delta RR$ are
significantly narrower [figure~\ref{fig4} (b)] and that they
are very well fitted to a normal distribution; however, the
mean values of the slopes differ from zero.

Because the dynamics of $RR$-intervals is evidently
non-stationary, we have applied detrended fluctuation analysis
(DFA) \cite{Peng:95} for estimation of the scaling exponents
$\beta$ for the apn{\oe}a sets $\{ RR_i\}^j$. In doing so, we
adapted the DFA method \cite{Peng:95} for short time-series and
used non-overlapped windows (see Appendix for details). Because
the time-series were short, time windows of length 4--15
$RR$-intervals were used to calculate $\beta$. For all measured
subjects, this procedure yielded values of $\beta$ lying within
the range $\beta \in(1.3:1.7)$, with a mean value of $1.45$. If
$RR$-intervals in the sets $\{ RR_i\}^j$ are replaced by
realizations of Brown noise (the integral of white noise)
keeping same  lengths of apn{\oe}a intervals, then the
calculation gives $\beta = 1.46 \pm 0.07$. Additionally, a
surrogate analysis was performed for each subject by random
shuffling of the time indices $i$ of $RR_i$-intervals, to
confirm the importance of time-ordering of the $RR$-intervals.
For each realization (set $\{ RR_i\}^j$), 100 surrogate sets
were generated, 100 values of $\beta$ were obtained, and the
mean value $\beta_s$ was calculated. Values of $\beta_s$ for
the surrogate sets lie in the same limits as those for the
original sets, but with a small bias between $\beta$,
calculated using original sets, and $\beta_s$ (see the Appendix
for values of $\beta$ and $\beta_s$). It means that one can see
a correlation between $RR$-intervals, but that it is weak.
Summarizing the DFA results, we can claim that the scaling
exponent $\beta$ is similar to that for free diffusion of a
Brownian particle, but there is nonetheless some correlation
between the $RR$-intervals.  We also applied aggregation
analysis \cite{West:05r} in a similar manner and arrived at
qualitatively the same conclusion. Note that in the contrast to
the initial idea of the DFA and aggregation analyses, which
were used for revealing long-range correlations in time series,
we have used these approaches to analyse the diffusion velocity
because they can cope with trends.  Long-range correlations
cannot be revealed in the described measurements.

\begin{figure}[t]
\centerline {\includegraphics[width=14cm]{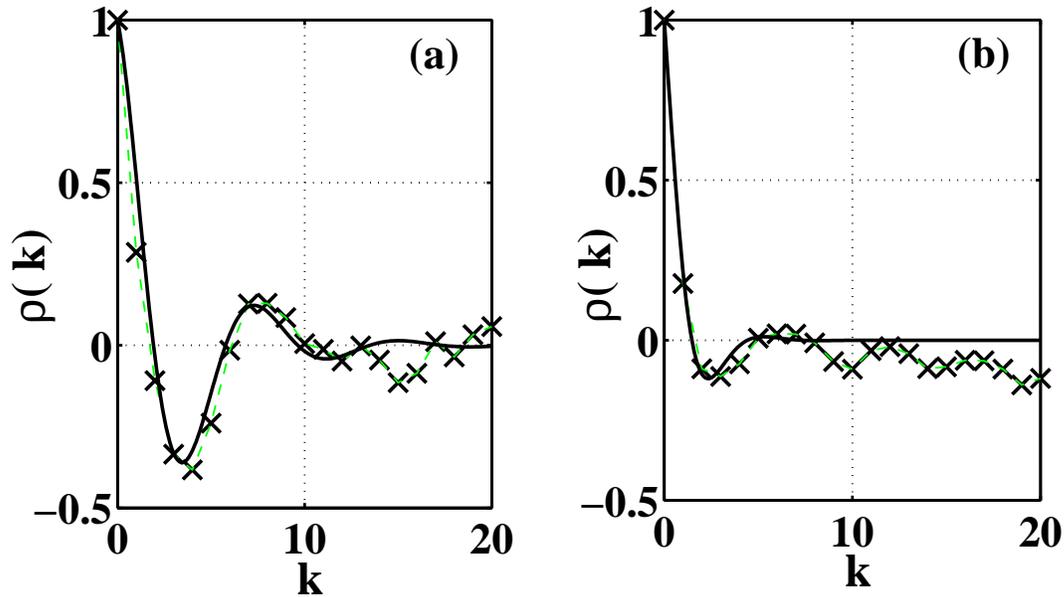} }
\caption{\label{fig5}  Examples of the autocorrelation function
$\rho ( k)$  (a) with  and (b) without an oscillatory component.
The crosses indicate $\rho( k )$ calculated on the basis of the
increments $\Delta RR$. The solid line corresponds to the
approximating curve $\rho^{a}( k )=\exp(-\gamma k )\cos(2\pi
\Omega k)$. }
\end{figure}

\begin{figure}[t]
\centerline {\includegraphics[width=14cm]{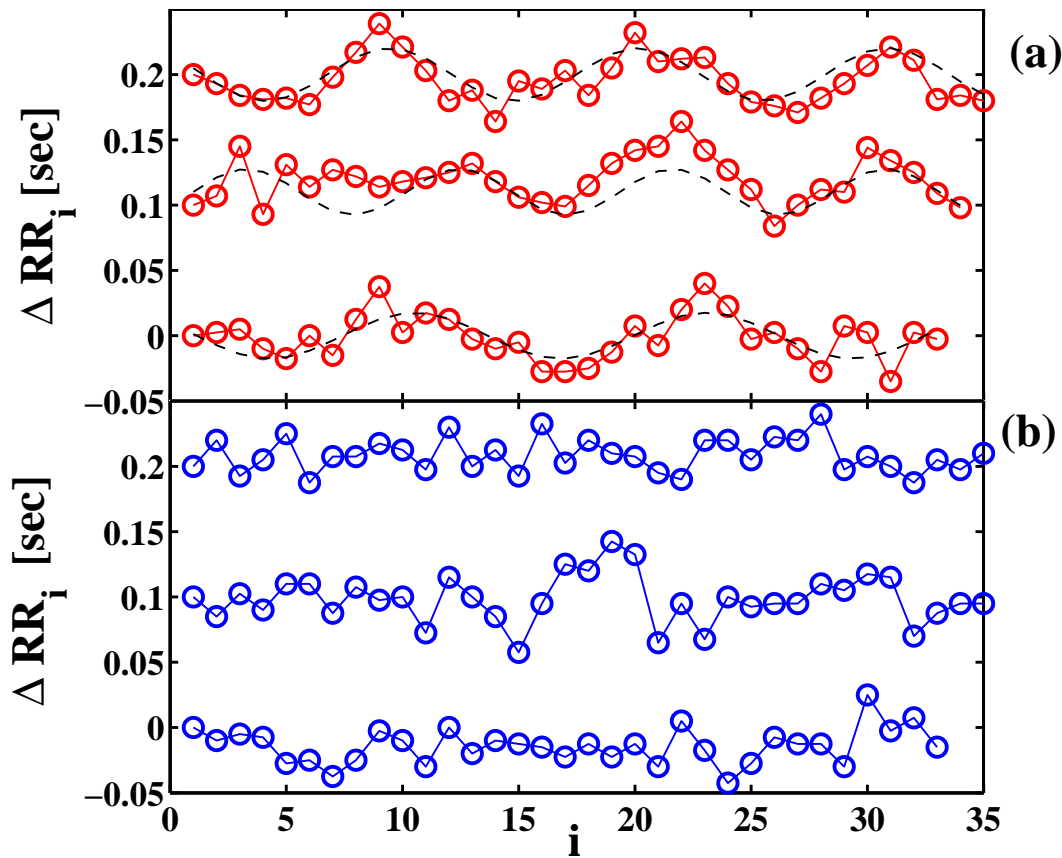} }
\caption{\label{fig5add}  Examples of apn{\oe}a intervals with
(a) and without (b) oscillation of HRV. The circles correspond
to the values of the increments $\Delta RR_i$ and the solid
lines connecting points are guides to the eye. The dashed lines
in figure (a) are added to reveal oscillations. The middle and
upper $\Delta RR_i$ time-series are shifted by 0.1 and 0.2
(sec) accordingly.
 }
\end{figure}

To estimate the strength of the correlation, {\it stationary}
time-series of the increments $\{\Delta RR_i\}^j$ were
considered. The autocorrelation function $\rho( k )$ was
calculated
\begin{eqnarray}
\label{acfRR}
\rho( k )=\frac{1}{(M-1)\sigma} \sum_{j=1}^{N}  \sum_{i=1}^{m^j-k}\widehat{RR}_i^j\widehat{RR}_{i-k}^j ; \\
M=\sum_{j=1}^{N} (m_j-k), \ \ \ \ \ \ \ \ \ \
\sigma=\frac{1}{M-1}\sum_{j=1}^{N} \sum_{i=1}^{m^j-k}
\left(\widehat{RR}_i^j\right)^2 . \nonumber
\end{eqnarray}
Here $\widehat{RR}_i^j=\Delta RR_i^j-\langle \Delta RR^j
\rangle$; the brackets $\langle \rangle$ denote calculation of
the mean value; $i$ and $j$ correspond to the heart beat number
and apn{\oe}a interval respectively, $k=0,1,\ldots$, $m^j$ is
the number of increments $\Delta RR$ in the $j^{\rm th}$
apn{\oe}a; $N$ is the total number of apn{\oe}a intervals.
Figure~\ref{fig5} presents examples of autocorrelation
functions. One of them has pronounced oscillations. An
approximation of $\rho( k)$ by the function $\rho^{a}( k
)=\exp(-\gamma k )\cos(2\pi \Omega  k)$ demonstrates that
oscillations occur with frequency near $0.1$ Hz, presumably
corresponding to myogenic processes \cite{Stefanovska:99a} or
(perhaps equivalently) to the Mayer wave associated with blood
pressure feedback \cite{Malpas:02,Julien:06}. Further
investigations via the parametrical spectral analysis for each
apn{\oe}a interval show that these oscillations are of an
on-off nature, i.e.\ observed for parts of the apn{\oe}a
intervals, and not in all of the measurements as can be seen in
figure~\ref{fig5} (b). Examples of apn{\oe}a intervals with and
without oscillations are shown in figure~\ref{fig5add}. When an
oscillatory component is present then its contribution to
$\rho( k )$ is much weaker than the contribution of the  noisy
component. The latter is characterized by a very short memory
as demonstrated by fast decay of $\rho( k )$.

\begin{figure}[t]
\noindent \includegraphics[width=7cm]{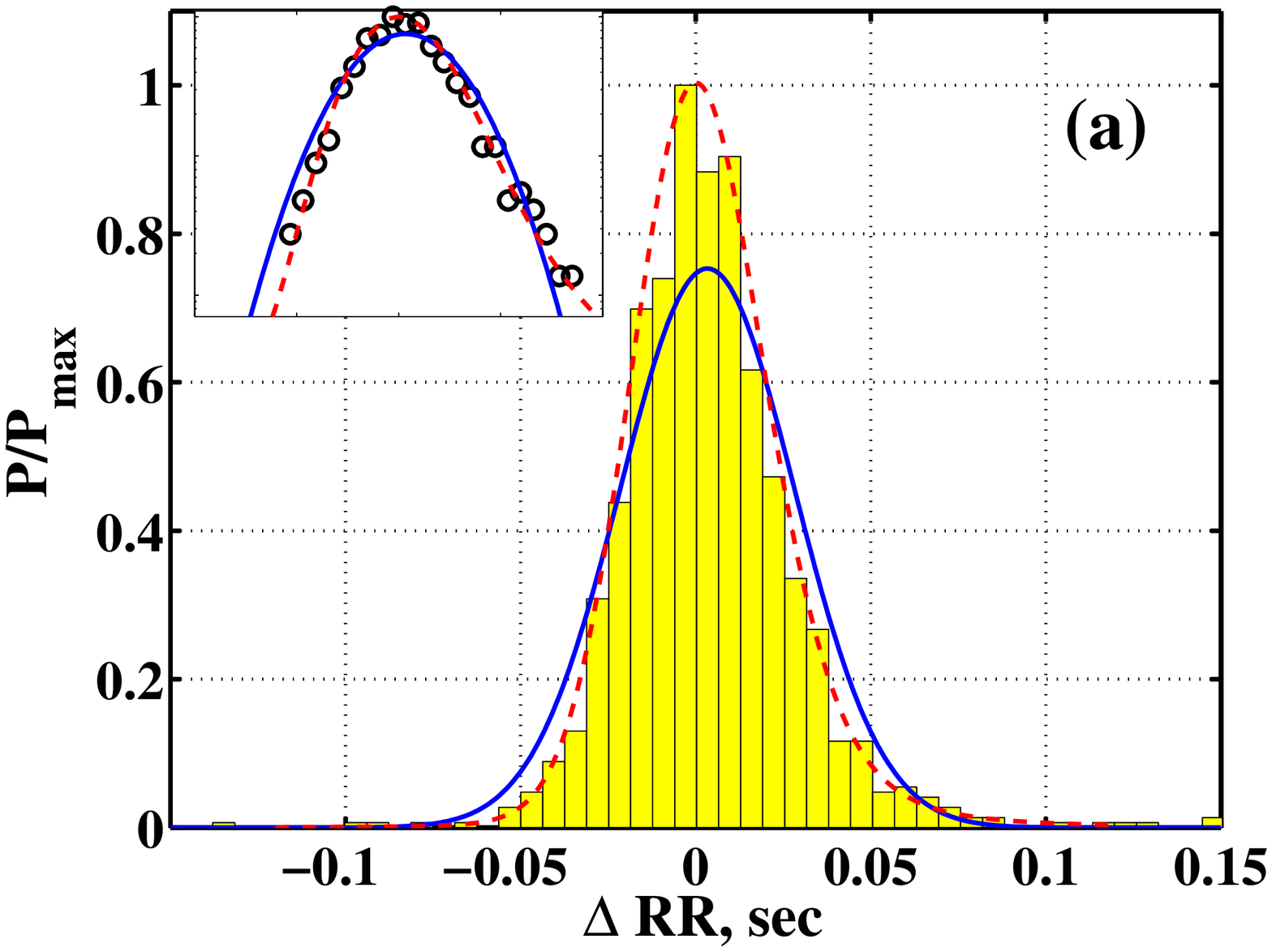}\ \ \ \ \ \ \includegraphics[width=7cm]{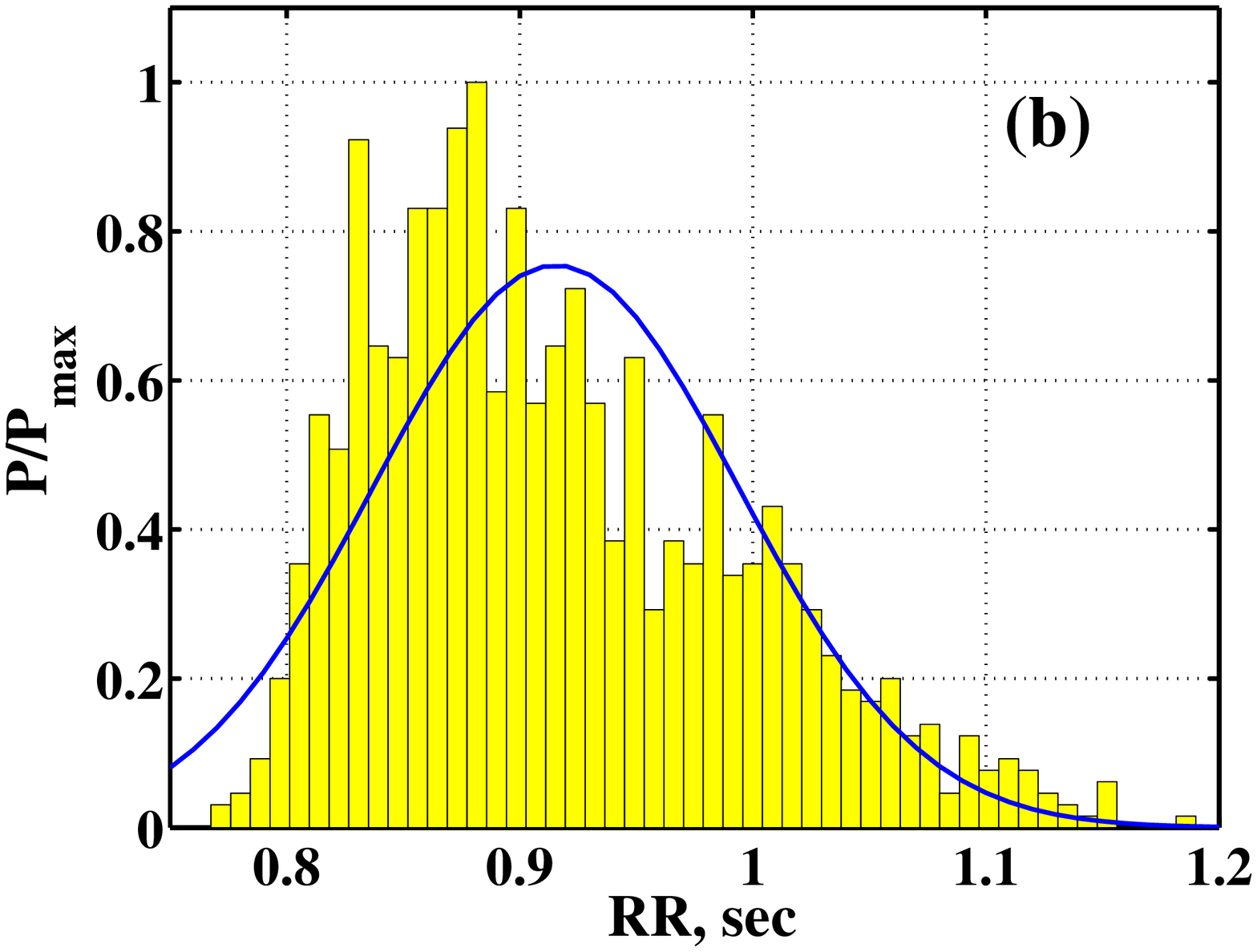}\\

\caption{\label{fig6} (Color online) Normalized probability
density functions (a) $P(\Delta RR)$ of increments of
$RR$-intervals and (b) $P(RR)$ of $RR$-intervals. In (a) the
full (blue) and dashed (red) lines are Gaussian and stable
distributions, respectively, fitted to the data. The insets
show the same distributions plotted with logarithmic ordinate
scales; the circles correspond to $P(\Delta RR)$. The stable
distribution in (a) is characterized by $\alpha=1.86$. In (b)
the full (blue) line is a Gaussian distribution fitted to the
data.}
\end{figure}

The properties of $\Delta RR$ can also be characterized by the
probability density function $P(\Delta RR)$ shown in
figure~\ref{fig6} (a). Figure (b) shows the probability density
function $P(RR)$ of RR-intervals for comparison. Following
\cite{Peng:93}, the $\alpha$-stable distribution has been
widely used to fit the distribution of increments $\Delta RR$,
and {\it strongly} non-Gaussian distributions were observed
\cite{Peng:93}. We perform a similar fitting applying special
software \cite{Nolan:98}. Since the distributions $P(\Delta
RR)$ are almost symmetrical, our attention was concentrated on
the tails, which were characterized by a stability index
$\alpha \in (0,2]$. The case of $\alpha=2$ corresponds to a
Gaussian and, if $\alpha <2$, the tails are wider than
Gaussian. Fitting to our results yields a stability index
$\alpha \in (1.8:2)$, and the goodness-of-fit test (modified
KS-test taking into account the weight to the tails
\cite{Nolan:98}) supports the fitting. Note that, although the
autocorrelation function $\rho( k )$ cannot be used for the
theoretical description of an $\alpha$-stable process
\cite{Samorodnitsky:94}, $\rho( k )$ is nonetheless applicable
for finite time-series.

If we consider the same length of realization using a Gaussian
random variable instead, we find $\alpha=1.99 \pm 0.01 $. It
means that the calculations of $\alpha$ are very robust. In
addition we carried out a stability test and it too supported
the fitting results. The obtained values of $\alpha \in
(1.8:2)$ differ significantly from the previously reported
values $\alpha \in (1.5:1.7)$ for 24h time-series of
$RR$-intervals \cite{Peng:93}.

Combining all the results, we conclude that the short-time
dynamics of $RR$-intervals can be described as a stochastic
process with stationary increments. This type of stochastic
processes was discussed by A. N. Kolmogorov
\cite{Kolmogorov:40} and applied to the description of a number
of different problems (see e.g.
\cite{Doob:90,Yaglom:62,Rytov:89} for further details). So, HRV
during apn{\oe}a interval cab be presented in the following
form
\begin{eqnarray}
\label{SPSI}
RR_i= RR_{i-1}  +\Delta RR_i,
\end{eqnarray}
where  $\Delta RR_i$ is a stationary discrete time stochastic
process. Note that the DFA calculation excludes a linear trend,
which is taken into account in Eq. (\ref{SPSI}) as non-zero
mean value of the increments, $\mu_j=\langle \Delta RR_i
\rangle_j$; in general case, $\mu_j$ is a random function of
$j$th apn{\oe}a interval. If one represents $RR$-intervals as a
sum of the linear trend and a random component:
\begin{eqnarray}
\label{SPLT}
RR_i= \mu_j \, i  +\xi_i,
\end{eqnarray}
then $\xi_i$ corresponds to the non-stationary process (\ref{SPSI}) with zero mean value of increments. In other words, the superimposed random component of HRV during apn{\oe}a intervals is described by   a non-stationary random process.

Increments $\Delta RR_i$ are characterized by a random
$\alpha$-stable process of short  memory, with a weak
intermittent oscillatory component of frequency $\sim$\,0.1~Hz.
In the zeroth approximation the increments can {\it safely} be
represented by an uncorrelated Gaussian random process but, in
the next approximation, a weak correlation must be included,
allowing for an intermittent oscillatory component, and for
weak non-Gaussianity of the distribution of increments $\Delta
RR$. These additions reveal, on the  one hand, that the
previously reported observation of a non-Gaussian distribution
of increments \cite{Peng:93} is a property of the intrinsic
heart rate regulatory system, but on the  other hand, that the
scaling ranges of the stability index $\alpha$ differ
significantly in the presence or absence of external
perturbations (including respiration) acting on the regulatory
system. Consequently an explanation of the scalings  reported
in \cite{Peng:93,Peng:95} should include analyses of the effect
of external perturbations and respiration, and not an analysis
of heart rate alone.

\section{Discussion}

\subsection{Non-stationarity of $RR$-intervals during apn{oe}a}

The results presented indicate that there is no firm set point
for the heart control system, and that the heart rhythm
exhibits diffusive behaviour. The slowest dynamics can be
described by a linear trend during apn{\oe}a intervals and its
presence can be treated as a slow regulatory/adaptation
component of the control system. The presence of the slow
time-scales is an established property of HRV \cite{Camm:96}
and their presence, even in the absence of the respiratory
perturbation, can be interpreted as an expected property.

On short time-scales of order several seconds, HRV shows a
diffusive dynamics too. It can be interpreted in two ways. One
possibility  is that  the control system does not firmly  trace
the base (slow) rhythm, because in  case of tracing, short
time-fluctuations should ``jump'' around the base rhythm and,
consequently, be stationary. Such a picture corresponds to zero
action of the control system if the heart rate is in a ``safe''
(for the whole cardiovascular system) interval, e.g. $RR \in
[RR_{low}:RR_{high}]$. Another possible explanation could be
that the control system is tracing the base rhythm but the
short-time fluctuations have a non-stationary character. It is
natural to expect that there could be other possible
explanations, and additional investigations are needed to reach
an understanding of the diffusive dynamics on short-time
scales.

In section 2 it is suggested that we should consider the
non-stationarity and diffusive dynamics of $RR$-intervals
within the framework of a stochastic process with independent
increments. It allows one to consider $RR$-intervals as
realizations of the so-called auto-regressive process that is
widely used in time-series analysis \cite{Box:94}.  It means
that the direct spectral estimation of $RR$-intervals,
currently used as one of the basic techniques \cite{Camm:96},
is not applicable here and that one must use the theory of
stochastic processes with stationary increments for  their
spectral decomposition \cite{Rytov:89}.  If in the presence of
respiration, the short-time stochastic component of HRV
preserves non-stationarity then spectral estimation based on
$RR$-intervals is not correct, and increments must be used
instead. Note, that the properties of short-time fluctuations
in the presence of respiration are far from being completely
understood.

\subsection{Non-Gaussianity and correlations of increments $\Delta RR$}

The theories of both stochastic processes with stationary
increments and of auto-regressive analysis place some
limitations on the analysed time-series. The first approach
requires the existence of finite second-order  momenta, whereas
the second approach assumes uncorrelated  statistics of
increments. Formally, however, non-Gaussianity of the
increments distribution means that the second-order momenta do
not exist \cite{Samorodnitsky:94}, but non-Gaussianity  can
still be incorporated into the auto-regressive description
\cite{Nikias:95}. And {\it vice versa}, the presence of
correlations in the increments dynamics requires a modification
of the standard auto-regressive approach, and it is one that
can be incorporated naturally into the general theory. In the
current investigation we ignore these issues. We calculate the
auto-correlation function and use model (\ref{SPSI}), because
the finite length of the time-series guarantees the existence
of the second-order momenta, and the simplicity of (\ref{SPSI})
means that the inclusion of the correlations is a trivial
extension.

Our consideration has the formal character of time-series
analysis because we do not incorporate any preliminary
information about the possible dynamics of $RR$-intervals. The
analysis is based on the use of a set of relatively short
time-series, a fact that defines our choice of simple
statistical measures. One cannot exclude the possibility that
the use of other approaches to such data might provide
additional insight into HRV dynamics. For example, the
fractional Brownian motion approach \cite{West:05r,Heneghan:00}
and the theory of discrete non-stationary non-Markov random
process \cite{Yulmetyev:02} represent different paradigms,
which are based on assumptions about the origin of the data.
Note, that despite a long history of developing the approaches
and their applications, the approaches of fractional Brownian
motion and of stable random process are not standardized tools,
whereas the approach of non-Markov random process is not so
popular. There is no definite recipe for choosing a set of
measures which can uniquely specify (or provide a good
description of) the properties of  a renewal (discrete time)
stochastic process.

\subsection{Modelling}

Another way of attempting to understand the results is to try
to reproduce the observed data properties from an appropriate
model. In the context of our experiments, the modelling should
consist of a simulation of the electrical activity of
sinoatrial node (SAN) where the heart beats are initiated. For
modelling, one option is to use a bottom-up approach, which is
currently a very popular technique within the framework of the
complexity paradigm. In fact, available SAN cellular models
allow one to incorporate many details of physiological
processes like the openings and closures of specific ion
channels \cite{Wilders:07}. However, despite the complexity of
the models (40--100 variables) many important features are
still missed. For example, the fundamentally stochastic
dynamics of ion channels is represented by equations that are
deterministic. Heterogeneity of the SAN cellular locations and
intercell communications are among other important open issues
\cite{Ponard:07,Dobrzynski:05}.

An alternative option is the top-down approach using
integrative phenomenological models. In contrast to detailed
cell models, a toy model of the heart as a whole unit can be
developed. It is known that an isolated heart, and a heart in
the case of a brain-dead patient \cite{Neumann:01} demonstrate
nearly periodic behaviour. So, it is reasonable to assume that
the observed HRV is induced by the neuronal heart control
system, which is a part of the central nervous system. The
control system includes a primary site for regulation located
in the medulla \cite{Klabunde:04}, consisting of a set of
neural networks with connections to the hypothalamus and the
cortex. The control is realized via two branches of the nervous
system: the parasympathetic (vagal) and the sympathetic
branches. Although many details of the control system are still
missing \cite{Armour:04,Doessel:06}, it is currently accepted
that the vagal branch operates on faster time scales than the
sympathetic one, and that each branch has a specific
co-operative action on the heart rate and the dynamics of SAN
cells.

Let us consider an integrate-and-fire (IF) model as a model of
a SAN cell in the leading pacemaker. These cells are
responsible for initiating the activity of SAN cells and,
consequently, that of the whole heart \cite{Dobrzynski:05}. The
dynamics of the IF model describes the membrane potential
$U(t)$ of the cell by the following equations
\begin{eqnarray}
\label{IFmodel}
\frac{dU}{dt} = \frac{1}{\tau} ~~~~~~~~~~~~~~~~~~~~~ & \mbox{if} \ \ \ U(t) < U_t \\
U(t)=U_r, \ \ t^*=t     & \mbox{if} \ \ \ U(t)=U_t \ \ \ \mbox{and} \ \ \ \frac{dU}{dt} > 0,
\end{eqnarray}
Here $1/\tau$ defines a slope of integration, $U_t$ is the
threshold potential, $U_r$ is the resting (hyperpolarization)
potential; the time $t^*$ corresponds to the cell firing, and
it is the difference between two successive firings that
determines the instantaneous heart period or $RR$-interval,
$RR_i=t^*_i-t^*_{i-1}$. It is known \cite{Klabunde:04} that
increasing sympathetic activity with a combination of
decreasing vagal activity leads to an increase in the heart
period, and {\it vice versa}. Direct stimulation of the
sympathetic branch leads to an increase of the integration
slope $1/\tau$ and a lowering of the threshold potential $U_t$,
whereas vagal activation has the opposite effects, and
additionally, lowers the resting potential $U_r$. Thus, the
neuronal activities can be taken into account as modulations of
the parameters of the IF model (\ref{IFmodel}). For reproducing
HRV during apn{\oe}a, therefore, it is enough to present any of
the parameters $\tau$, $U_t$ or $U_r$ as a stochastic variable
of the form (\ref{SPSI}), for example,
$U_t(t^*_i)=U_t(t^*_{i-1})+\xi_i$, where $\xi_i$ are random
numbers having the stable distribution.

However, the use of more realistic (than IF) models with
oscillatory dynamics, for example Fitzhugh-Nagumo
\cite{FitzHugh:61} or Morris-Lecar \cite{Morris:81} models,
makes the reproduction of the experimental results a more
difficult but interesting task. Currently it is unclear whether
it is possible to obtain a stable distribution of increments by
consideration of the Gaussian type of fluctuations alone, or
whether one should use fluctuations characterizing by a stable
distribution. This point demands further investigation.

\section{Conclusion}

In summary, our experimental modification of the respiration
process reveals that the intrinsic dynamics of the heart rate
regulatory system exhibits stochastic features and can be
viewed as the integrated action of many weakly interacting
components. Even on a short time scale (less then half a
minute) the heart rate is non-stationary and exhibits diffusive
dynamics with superimposed intermittent $\sim$\,0.1~Hz
oscillations.  The intrinsic dynamics can be described as a
stochastic process with independent increments and can be
understood within the framework of many-body dynamics as used
in statistical physics. The large number of independent
regulatory perturbations produce a noisy regulatory background,
so that the dynamics of the regulatory rhythm is close to
classical Brownian motion. However there are indications of
non-Gaussianity of increments and weak but important
correlations on short time-scales. The reproduction of these
features, especially the non-Gaussianity property, is an open
problem even in simple toy models.

These results are important both for understanding the general
principles of regulation in biological systems, and for
modeling cardiovascular dynamics. Furthermore, the results
presented may possibly lead to a new clinical classification of
states of the cardiovascular system by analysing the intrinsic
dynamics of the heart control system as suggested in
\cite{Doessel:06}.

\ack

The research was supported by the Engineering and Physical Sciences
Research Council (UK) and by the Wellcome Trust.

\appendix
\setcounter{section}{1}

\section*{Appendix}

Some details of the measurements and calculations are
summarized in this section.

The ECG was measured by standard limb (Einthoven) leads and the
respiration signal was measured by a thoracic strain gauge
transducer. The signals were digitized by a 16-bit
analog-to-digital converter with a sampling rate of 2 kHz. The
ECG and respiration signals were recorded  over 45-60 minutes
and time locations of $R$-peaks in the ECG signals were defined
and time intervals between two subsequent $R$-peaks (the so
called $RR$-intervals) are used to form HRV signal.

Respiration-free intervals were produced by the {\it
intermittent} respiration, involving an alternation between
normal  breaths and apn{\oe}a intervals. The durations of both
normal breaths and apn{\oe}a intervals were fixed at 30 sec.
The respiration signal was used to identify apn{\oe}a
intervals. Finally, the set of time-series of $RR$-interval
$\{RR_i\}^j$ was formed for each subject; here $i=1,2\ldots$
labels the $i$th  $RR$-intervals, and $j=1,2\ldots$ labels the
$j$th interval of apn{\oe}a. For each interval of apn{oe}a,
time series of the differential increments $\Delta
RR_i=RR_{i+1}-RR_i$ were produced and they also form a set $\{
\Delta RR_i\}^j$ for each subject. The number of $RR$-intervals
in each apn{\oe}a interval is different, depending on the heart
rate of the subject. The total number of apn{\oe}a intervals
also differ for each subject. The mean heart rate $\langle RR
\rangle$ during apn{\oe}a intervals and the total number $J$ of
intervals for each measured subject are presented in table
\ref{t1}.

\begin{table}
\caption{\label{t1} Data for each subject. $\langle RR \rangle$
is the mean heart rate during apn{\oe}a. $J$ is the total
number of apn{\oe}a intervals. $\beta$ is the DFA scaling
exponent calculated for the apn{\oe}a set $\{ RR_i\}^j$.
$\beta_s$ is the mean value of the DFA scaling exponent
calculated for surrogate data, which were generated by random
shuffling of the time indices $i$ of $RR_i$-intervals. $b$ is
the   scaling exponent of the aggregation analysis. $\gamma$
and $\Omega$ correspond to the values of parameters for the
function $\rho^{a}( k )=\exp(-\gamma k )\cos(2\pi \Omega  k)$
which approximates the autocorrelation function $\rho(k)$.
$\alpha$ is the stability index of the distribution $P(\Delta
RR)$.}
\begin{indented}
\item[]\begin{tabular}{@{}llllllllll}
\br
Subject & $\langle RR \rangle  $ (sec) &  $J$~~~~  &  $\beta$~~~~&  $\beta_{s}$~~~~ & $b$~~~~ & $\gamma$~~~~ & $\Omega$~~~~ & $\alpha$~~~~ \\
\mr
S1 & 1.01  & 45   & 1.39 & 1.47 & 1.86 & 0.81 & 0.17 & 1.83 \\
S2 & 0.77 &  46  &  1.46 & 1.44 & 1.83 & 0.21 & 0.09 & 1.95 \\
S3 & 1.10 &  47  &  1.43 & 1.53 & 1.96 & 1.01 & 0.22 & 1.79 \\
S4 & 0.75 &  47  &  1.58 & 1.60 & 1.91 & 0.15 & 0.08 & 1.90 \\
S5 & 0.91  &  60  & 1.42 & 1.48 & 1.82 & 0.28 & 0.13 & 1.86 \\
\br
\end{tabular}
\end{indented}
\end{table}

For the application of the DFA and aggregation analyses we
adapted the approaches described in \cite{Peng:95} and
\cite{West:05r}, respectively, to treat the available sets of
short time series $\{ RR_i\}^j$.

\begin{figure}[t]
\noindent  \includegraphics[width=7 cm]{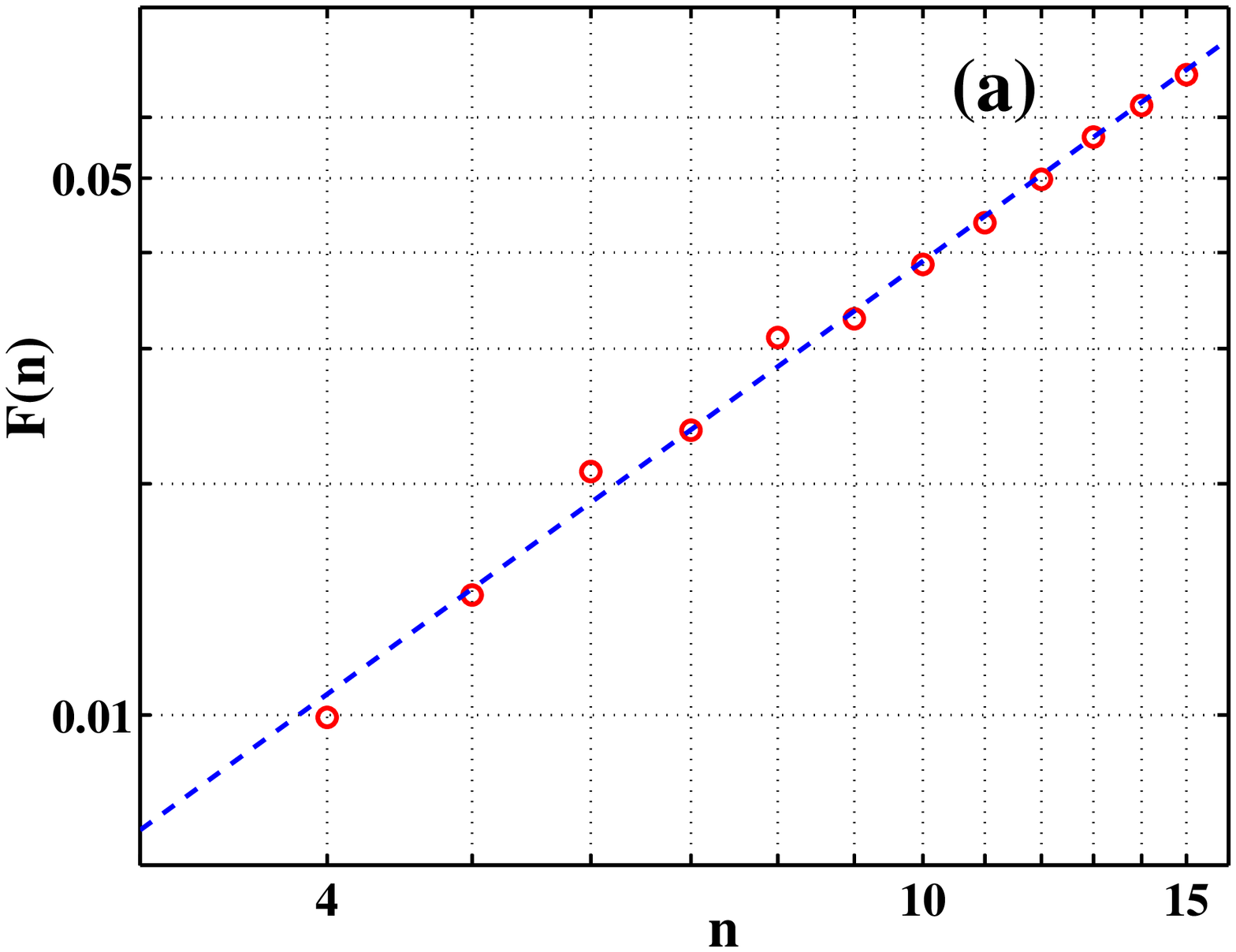} ~~~~~~\includegraphics[width=7 cm]{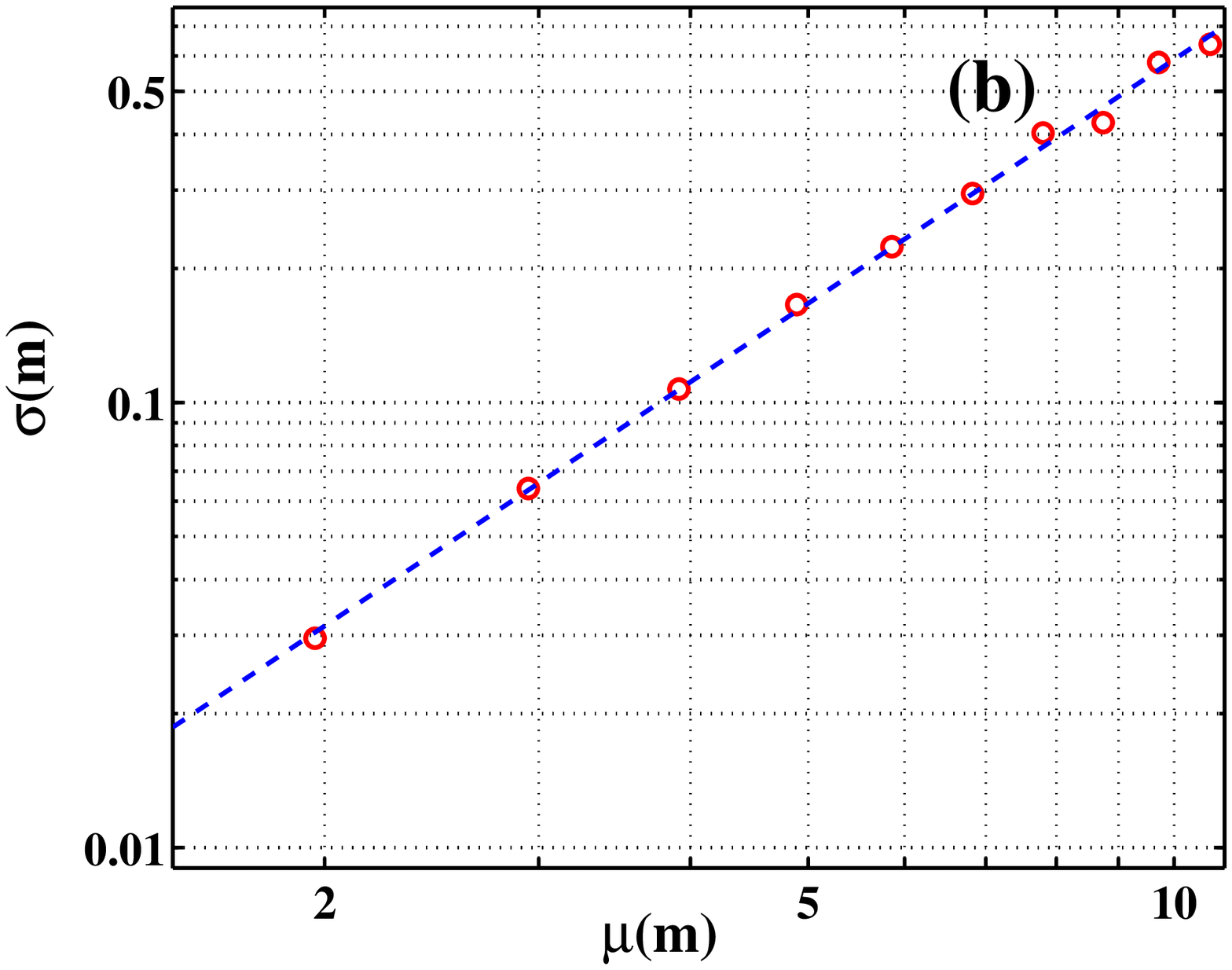}
\caption{\label{figa1} (a) The scaling function $F(n)$ (circles) and its approximation (dashed line) by $F(n) \propto n^\beta$ ($\beta=1.39$) are shown. (b) The dependence (circles) of the variance $\sigma(m)$ on the mean value $\mu(m)$ for $m=1,\ldots 10$ and its approximation (dashed line) by
$\sigma (m) \propto \mu(m) ^b$ ($b=1.82$) are shown.}
\end{figure}

The DFA exponent $\beta$ was calculated in the following way.
First, the initial set $\{ RR_i\}^j$ was transformed to another
set $\{ y(k) \}^j$ by the following expression:
\begin{equation}
\label{cumsum}
y(k)=\sum_{i=1}^k RR_i,
\end{equation}
where $k=1\ldots M_j$ and $M_j$ is the number of $RR$-intervals
for $j$th apn{oe}a interval. For each length $n=4,\ldots 15$ of
time window a set of linear trends $\{ y^n(k) \}^j$ was
calculated (see \cite{Peng:95} for details), where $y^n(k)=k
\cdot a_m^n +b_m^n$,  $m=1,\ldots \lfloor {M_j/n} \rfloor$,
$\lfloor x \rfloor=\max \{n \in {\mathbb Z}| n\le x \}$ is the
floor function of $x$. Then a set of scaling function $\{
\tilde{F}(n) \}^j$ was calculated for each value of $n$ by use
of the expression
\begin{equation}
\label{psc}
\tilde{F}(n)=\sum_k^{N_j} [y(k)-y^n(k)]^2,
\end{equation}
where $N_j=\lfloor {M_j/n} \rfloor \cdot n$. Further the
scaling functions $F(n)$ were calculated as
\begin{equation}
F(n)=\sqrt{\frac{1}{N-1} \sum_{j=1}^{J} \tilde{F}_j(n)},
\end{equation}
where $J$ is the number of apn{oe}a intervals for the given
subject, $N=\sum_{j=1}^{J} N_j$. Finally, the scaling exponent
$\beta$ was determined as a slope of the function $\log [F(n)]
\propto \beta \log (n)$ (see figure \ref{figa1} (a)). The
values of $\beta$ for the different subjects are shown in table
\ref{t1}.

The aggregation analysis consists of three steps and the final
result is the scaling exponent $b$. The first step is the
creation of a set of aggregated time series $\{ z_m(k) \}^j$
for different $m=1,\ldots 10$:
\begin{equation}
\label{ats}
z_m(k)=\sum_{i=k}^{k+m} RR_i,
\end{equation}
where $k=1,\ldots M_j-m$. Then a realization $z_m(k)$ was
formed from the set $\{ z_m(k) \}^j$: $z_m(k)=\{ z_m(k) \}^j=\{
z_m(k) \}^1,\ldots\{ z_m(k) \}^{J}$. The second step includes
the calculation of the mean value $\mu(m)$ and variance
$\sigma(m)$ of the time-series $z_m(k)$:
\begin{equation}
\label{ss}
\mu(m) = \frac{1}{M} \sum_{k=1}^{M} z_m(k), \ \ \ \ \sigma(m) = \frac{1}{M-1} \sum_{k=1}^{M} \left[ z_m(k) -\mu(m) \right]^2,
\end{equation}
where $M$ is the whole length of time series $z_m(k)$. The
slope $b$ of the function  $\log [\sigma (m)] \propto b \log
[\mu (m) ]$  was calculated in the third step (see figure
\ref{figa1} (b)). The values of $b$ for each subject are shown
in table \ref{t1}.

To verify the robustness of the calculations of exponents
$\beta$ and $b$ we have performed calculations with the same
number of RR-intervals as well as  the same structure of
apn{\oe}a intervals but by using realizations of Brown noise
generated by computer. In other words, in the procedures
described above we replaced $\{ RR_i \}^j$ by $\{W_i \}^j$,
where $W_i=W_{i-1}+0.2\cdot \xi_i$ for $i=2,\cdots M_j$,
$W_1=RR_1$, and $\xi_i$ are random numbers having the normal
distribution with mean zero value and unit variance; the
numbers $\xi_i$ are different for different $j$th intervals of
apn{\oe}a. We performed 100 calculation of $\beta$ and $b$ for
different sets $\{W_i \}^j$ for each subject. Theoretical
values of $\beta$ and $b$ for the Brown noise are $\beta=1.5$
and $b=2$ correspondingly. The calculations with Brown noise
gave $\beta = 1.46 \pm 0.07$ and $b=1.84\pm 0.04$. Here data
were merged for all subjects and are presented in the form of a
mean value $\pm$ its standard deviation. It means that there is
a systematic error related to the length and data structure, a
general error of calculation in respect to the theoretical
values  for $\beta$  is $~1.5$ and for $b$  is $~2.0$. However
the standard deviations of the calculated values are rather
small and, consequently, we can conclude that our calculations
of $\beta$ and $b$ are robust.

\section*{References}


\begin{thebibliography}{10}
\expandafter\ifx\csname url\endcsname\relax
  \def\url#1{{\tt #1}}\fi
\expandafter\ifx\csname urlprefix\endcsname\relax\def\urlprefix{URL }\fi
\providecommand{\eprint}[2][]{\url{#2}}

\bibitem{Ott:93}
Ott E 1993 {\em Chaos in Dynamical Systems\/} (Cambridge, UK: Cambridge
  University Press)

\bibitem{Frisch:95}
Frisch U 1995 {\em Turbulence: The Legacy of A.N. Kolmogorov\/} (Cambridge, UK:
  Cambridge University Press)

\bibitem{Mandelbrot:82}
Mandelbrot B~B 1982 {\em The Fractal Geometry of Nature\/} ({W. H. Freeman})

\bibitem{Jensen:98}
Jensen H~J 1998 {\em Self-Organized Criticality: Emergent Complex Behavior in
  Physical and Biological Systems\/} (Cambridge, UK: {Cambridge University
  Press})

\bibitem{Kurths:95}
Kurths J, Voss A, Saparin P, Witt A, Kleiner H~J and Wessel N 1995 {\em
  Chaos\/} {\bf 5}(1) 88--94

\bibitem{Pincus:91}
Pincus S~M {1991} {\em {Proc.\ Nat.\ Acad.\ Sci.}\/} {\bf {88}}({6})
  {2297--2301}

\bibitem{Costa:02}
Costa M, Goldberger A~L and Peng C~K 2002 {\em Phys. Rev. Lett.\/} {\bf 89}(6)
  068102

\bibitem{Raab:06}
Raab C, Wessel N, Schirdewan A and Kurths J 2006 {\em Phys.\ Rev.\ E\/} {\bf
  73}(4) 041907

\bibitem{Kiyono:05}
Kiyono K, Struzik Z~R, Aoyagi N, Togo F and Yamamoto Y 2005 {\em Phys.\ Rev.\
  Lett.\/} {\bf 95}(5) 058101

\bibitem{Peng:95}
Peng C~K, Havlin S, Stanley H~E and Goldberger A~L 1995 {\em Chaos\/} {\bf
  5}(1) 82--87

\bibitem{Hausdorff:96}
Hausdorff J~M and Peng C~K 1996 {\em Phys. Rev. E\/} {\bf 54}(2) 2154--2157

\bibitem{Pilgram:99}
Pilgram B and Kaplan D~T 1999 {\em Amer.\ J.\ of Physiol.\ -- Regulatory,
  Integrative and Comparative Physiol\/} {\bf 276}(1) R1--R9

\bibitem{Costa:05}
Costa M, Goldberger A~L and Peng C~K 2005 {\em Phys. Rev. Lett.\/} {\bf 95}
  198102

\bibitem{Ivanov:99}
Ivanov P~C, Amaral L~A~N, Goldberger A~L, Havlin S, Rosenblum M~G, Struzik Z~R
  and Stanley H~E 1999 {\em Nature\/} {\bf 399}(6735) 461--465

\bibitem{Ivanov:01}
Ivanov P~C, Amaral L~A~N, Goldberger A~L, Havlin S, M~G~Rosenblum H~E~S and
  Struzik Z~R 2001 {\em Chaos\/} {\bf 11}(3) 641--652

\bibitem{Ivanov:98b}
Ivanov P~C, Amaral L~A~N, Goldberger A~L and Stanley H~E 1998 {\em Europhys.\
  Lett.\/} {\bf 43}(4) 363--368

\bibitem{Stefanovska:99a}
Stefanovska A and Bra{\v{c}}i{\v{c}} M 1999 {\em Contemporary Phys.\/} {\bf
  40}(1) 31--55

\bibitem{Eckberg:03}
Eckberg D~L 2003 {\em J.\ Physiol.\ (Lond.)\/} {\bf 548}(2) 339--352

\bibitem{West:05r}
West B~J, Griffin L~A, Frederick H~J and Moon R~E {2005} {\em {Resp.\ Physiol.\
  and Neurobiol.}\/} {\bf {145}}({2-3}) {219--233}

\bibitem{Doessel:06}
Doessel O, Reumann M, Seemann G and Weiss D {2006} {\em {Biomedizinische
  Technik}\/} {\bf {51}}({4}) {205--209}

\bibitem{Parkes:06}
Parkes M~J {2006} {\em {Exper.\ Physiol.}\/} {\bf {91}}({1}) {1--15}

\bibitem{Malpas:02}
Malpas S~C 2002 {\em Am.\ J.\ Physiol.: Heart.\ Circ.\ Physiol.\/} {\bf 282}
  H6--H20

\bibitem{Julien:06}
Julien C {2006} {\em {J.\ Appl.\ Physiol.}\/} {\bf {101}}({2}) {684}

\bibitem{Peng:93}
Peng C~K, Mietus J, Hausdorff J~M, Havlin S, Stanley H~E and Goldberger A~L
  1993 {\em Phys.\ Rev.\ Lett.\/} {\bf 70}(9) 1343--1346

\bibitem{Nolan:98}
Nolan J~P {1998} {\em {Stat.\ and Probabil.\ Lett.}\/} {\bf {38}}({2})
  {187--195}

\bibitem{Samorodnitsky:94}
Samorodnitsky G and Taqqu M~S 1994 {\em Stable Non-Gaussian Random Processes:
  Stochastic Models with Infinite Variance\/} (London, UK: Chapman and Hall)

\bibitem{Kolmogorov:40}
Kolmogorov A~N 1940 {\em Doklady Akademii Nauk SSSR\/} {\bf 26}(1) 6--9

\bibitem{Doob:90}
Doob J~L 1990 {\em Stochastic Processes\/} (New York: Wiley-Interscience)

\bibitem{Yaglom:62}
Yaglom A~M 1962 {\em An Introduction to the Theory of Stationary Random
  Functions\/} (Englewood Cliffs, NJ.: Prentice-Hall)

\bibitem{Rytov:89}
Rytov S~M, Kravtsov Y~A and Tatarskii V~I 1988 {\em Principles of Statistical
  Radiophysics\/} vol 2: Correlation Theory of Random Processes
  (Springer-Verlag)

\bibitem{Camm:96}
Camm A~J, Malik M, Bigger J~T and {\em et al} 1996 {\em Circulation\/} {\bf
  93}(5) 1043--1065

\bibitem{Box:94}
Box G, Jenkins G~M and Reinsel G 1994 {\em Time Series Analysis: Forecasting \&
  Control (3rd Edn)\/} (Englewood Cliffs, NJ: Prentice Hall)

\bibitem{Nikias:95}
Nikias C~L and {Min Shao} 1995 {\em Signal Processing with Alpha-stable
  Distributions and Applications\/} (New York, NY, USA: Wiley-Interscience)

\bibitem{Heneghan:00}
Heneghan C and McDarby G 2000 {\em Phys.\ Rev.\ E\/} {\bf 62}(5) 6103--6110

\bibitem{Yulmetyev:02}
Yulmetyev R, H\"anggi P and Gafarov F 2002 {\em Phys. Rev. E\/} {\bf 65}(4)
  046107

\bibitem{Wilders:07}
Wilders R {2007} {\em {Med.\ and Biol.\ Eng.\ and Comp.}\/} {\bf {45}}({2})
  {189--207}

\bibitem{Ponard:07}
Ponard J~G~C, Kondratyev A~A and Kucera J~P {2007} {\em {Biophys.\ J.}\/} {\bf
  {92}}({10}) {3734--3752}

\bibitem{Dobrzynski:05}
Dobrzynski H, Li J, Tellez J, Greener I~D, Nikolski V~P, Wright S~E, Parson
  S~H, Jones S~A, Lancaster M~K, Yamamoto M, Honjo H, Takagishi Y, Kodama I,
  Efimov I~R, Billeter R and Boyett M~R {2005} {\em {Circulation}\/} {\bf
  {111}}({7}) {846--854}

\bibitem{Neumann:01}
Neumann T, Post H, Ganz R~E, Walz M~K, Skyschally A, Schulz R and Heusch G 2001
  {\em Zeitschrift f\"{u}r Kardiologie\/} {\bf 90}(7) 484--491

\bibitem{Klabunde:04}
Klabunde R~E 2004 {\em Cardiovascular Physiology Concepts\/} (Philadelphia,
  USA: Lippincott Williams \& Wilkins)

\bibitem{Armour:04}
Armour J~A {2004} {\em Amer.\ J.\ Physiol.\ -- Regulatory Integrative and
  Comparative Physiol.\/} {\bf {287}}({2}) {R262--R271}

\bibitem{FitzHugh:61}
Fitzhugh R {1961} {\em {Biophys.\ J.}\/} {\bf {1}}({6}) {445}

\bibitem{Morris:81}
Morris C and Lecar H {1981} {\em {Biophys.\ J.}\/} {\bf {35}}({1}) {193--213}

\end{thebibliography}

\providecommand{\newblock}{}

\end{document}